\documentclass[12pt]{article}
\usepackage{pdproc}
\usepackage{epsfig}

 \newcommand{\numu}{\ensuremath{\nu_\mu }}
\newcommand{\nue}{\ensuremath{\nu_e}}
\newcommand{\nutau}{\ensuremath{\nu_\tau}}

\newcommand{\numunue}{\ensuremath{\numu \rightarrow \nue}}
\newcommand{\nubarmu}{\ensuremath{\overline{\nu}_\mu }}

\newcommand{\pnumunue}{\ensuremath{P(\numu \rightarrow \nue)}}

\newcommand{\dmatm}{\ensuremath{\delta m^2_{atm}}}

\newcommand{\thetaot}{\ensuremath{\theta_{13}}}
\newcommand{\thetatt}{\ensuremath{\theta_{23}}}

\newcommand{\sthetatt}{\ensuremath{{\rm sin}^2(2\theta_{23})}}
\newcommand{\nubare}{\ensuremath{\overline{\nu}_{e} }}

\begin{document}

\renewcommand{\thefootnote}{\alph{footnote}}

\title{
 PHYSICS POTENTIAL OF VERY INTENSE CONVENTIONAL NEUTRINO BEAMS}

\author{ JUAN JOS\'E G\'OMEZ CADENAS}

\address{ Instituto de F\'{\i}sica Corpuscular, IFIC \\
Edificio de Institutos de Paterna, 46071 Valencia, Spain \\
and CERN \\
 Ch-1211 Geneve 23, Switzerland\\
 {\rm E-mail: gomez@mail.cern.ch, gomez@ific.uv.es}}
  \centerline{\footnotesize as a rapporteur from the CERN working group
  on Super Beams}
\author{A. BLONDEL$^1$, J. BURGUET-CASTELL$^2$, D. CASPER$^3$, M. DONEGA$^1$, S. GILARDONI$^1$,
P. HERN\'ANDEZ$^4$ and M. MEZZETTO$^5$}

\address{(1) D\'epartement de Physique, Universit\'e de Geneve, Switzerland \\
(2) IFIC, Edificio de Institutos de Paterna, Paterna, Valencia,
Spain \\
(3) University of California at Irvine, USA \\
(4) INFN, sezione di Padova, Italy \\
(5) CERN, Ch-1211 Geneve 23, Switzerland}

\abstract{The physics potential of high intensity conventional
beams is explored. We consider a low energy super beam which could
be produced by a proposed new accelerator at CERN, the Super
Proton Linac. Water Cherenkov and liquid oil scintillator
detectors are studied as possible candidates for a neutrino
oscillation experiment which could improve our current knowledge
of the atmospheric parameters \dmatm,\thetatt~ and measure or
severely constrain the parameter connecting the atmospheric and
solar realms, \thetaot. It is also shown that a very large water
detector could eventually observe leptonic CP violation. The reach
of such an experiment to the neutrino mixing parameters would lie
in-between the next generation of neutrino experiments (MINOS,
OPERA, etc) and a future neutrino factory.}

\normalsize\baselineskip=15pt

\section{Introduction}
The notion of ``super beams" was introduced by B. Richter, who
suggested\cite{Richter} that a conventional neutrino beam of very
high intensity could be competitive with the pure two-flavor beams
produced by the neutrino factory. Recent
work\cite{barger1,barger2} has considered in great detail the
potential of generic super beams, with neutrino energies ranging
from 1 to 50 GeV and baselines spanning from 200 to 7000
kilometers. A large variety of detector technologies, including a
liquid argon TPC, a fine grain iron calorimeter and water
Cherenkov detectors \'a la Super-Kamiokande have been discussed as
potential candidates for a super beam experiment. The general
conclusion reached in\cite{barger1,barger2} is that super beams
can largely improve in our knowledge of \dmatm,\thetatt~ and
\thetaot, as well as providing some sensitivity to a CP violating
phase $\delta$, if the solution of the solar neutrino problem lies
in the upper region of the Large Mixing Angle (LMA-MSW). On the
other hand, it is also concluded that ultimate sensitivity to the
above parameters, in particular to $\delta$, will require the pure
and intense beams of a neutrino factory.

In this paper we present a complementary approach to the work
referred above. We consider only a super beam of very low neutrino
energy, 250 MeV on average, which was not studied
in\cite{barger1,barger2}. Such a beam will be produced by the very
intense Super Proton Linac\cite{spl}, a future facility planned at
CERN. Furthermore, we restrict ourselves to those technologies
which afford truly massive targets (a must, given the low energy
of our neutrinos), and therefore consider only water Cherenkov and
liquid scintillating oil detectors. In order to estimate the
experimental response (e.g., signal efficiency as well as beam and
detector-induced backgrounds) we have performed for the water
Cherenkov detector a full simulation followed by a detailed
analysis using the Super-Kamiokande tools, in contrast with the
simple estimations made in\cite{barger1,barger2} and, indeed,with
our own educated guesses for the liquid scintillating oil
detector.

This paper is organized as follows. In section 2 we briefly
address the general features of super beams. In section 3 the
Super Proton Linac (SPL hereafter) and the resulting, low-energy
neutrino beam are described. In section 4 we discuss our detector
scenarios. An estimation of the sensitivity to the atmospheric
parameters \dmatm,\thetatt, as well as to \thetaot~ and the CP
violating phase $\delta$ is presented in section 5. In section 6
we conclude.

\section{Conventional super beams}
\label{sec:beams}

A conventional neutrino beam is produced by hitting a nuclear
target with an intense hadron beam, then sign-selecting and
letting decay the resulting hadrons through a beam decay tunnel.
At the end of the tunnel there is an absorber, where the copiously
produced muons (a by product of pion and kaon decay) are ranged
out before most of them can decay.

\begin{figure}[htb]
\centerline{\epsfig{file=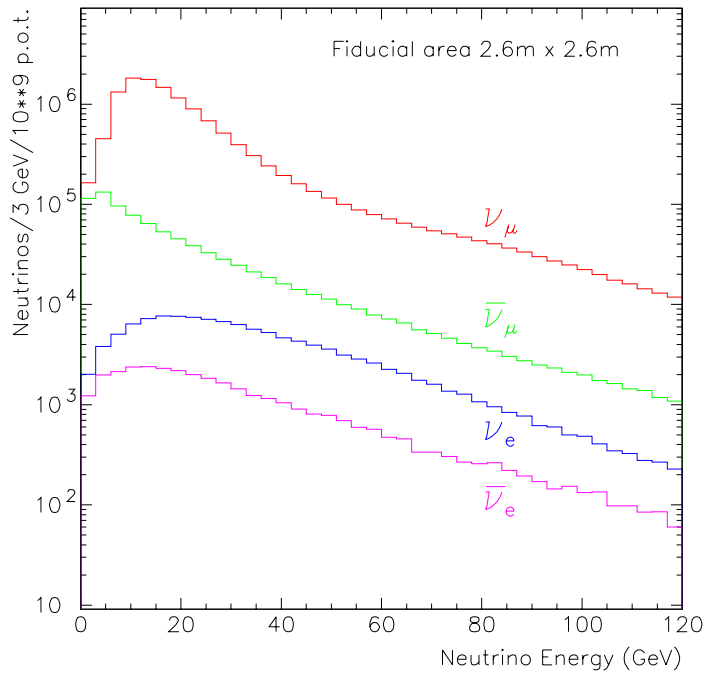,width=10.0cm}}
\caption{Fluxes produced by the SPS neutrino beam of the CERN West
area. Notice that the beam is mostly made of \numu~ but there are
contaminations of all other neutrino species, except \nutau.}
\label{fig:nomad}
\end{figure}

The resulting neutrino beam is mostly made of \numu (assuming that
$\pi^+$ were selected). Nevertheless, kaon and muon decays result
in small but sizeable contaminations of \nue~ and \nubare.
Opposite sign pion feed-through yields also some contamination of
\nubarmu. As an example, Figure \ref{fig:nomad} shows the neutrino
beam spectra produced by the 450 GeV Super Proton Syncroton, at
CERN, which illuminated the NOMAD\cite{nomad} and
CHORUS\cite{chorus} experiments.

Notice that the contamination of other neutrino species is a
handicap for the so-called neutrino oscillation {\it appearance}
experiments, in which one searches for a flavor not originally in
the beam. It has been stressed\cite{adr}that the best way of
measuring \thetaot~ and $\delta$ is through the transitions
$\nu_\mu \rightarrow \nu_e$. Any contamination of \nue~ in the
original beam must be subtracted, resulting in loss of
sensitivity.

\begin{figure}[htb]
\centerline{\epsfig{file=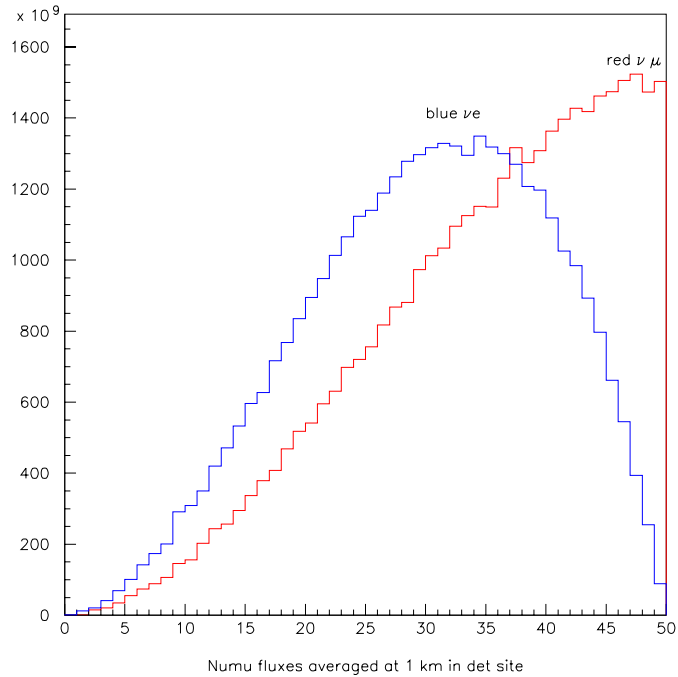,width=10.0cm}}
\caption{Neutrino beams produced by positive muon decay in an
accumulator ring. Notice that this is a pure two-flavor beam,
which no contamination of other neutrino species.}
\label{fig:nufact}
\end{figure}

Indeed, this is the key advantage of the neutrino factory beams,
produced by the decay of muons circulating in an storage ring,
over conventional beams. As shown in Figure \ref{fig:nufact}, the
decay of (say) positive muons result in a beam of {\it pure} \nue~
and \nubarmu, thus, there is no beam contamination (assuming, of
course, that one is able to measure the charge of the produced
lepton) to transitions of the type $\nu_e \rightarrow \nu_\mu$.

A super beam is nothing but a conventional beam of stupendous
intensity. Thus,(for $\pi^+$ selected in the horn), its basic
composition is \numu~ with small admixtures of \nue, \nubare~ and
\nubarmu. To gain some appreciation of the relative sensitivity of
a conventional neutrino beam and a neutrino factory beam, it is
useful to estimate the sensitivity to a $ \nu_\mu \rightarrow
\nu_e $ oscillation search in the appearance mode, {\it assuming a
perfect detector}. Let us assume that the product of the neutrino
beam intensity, running time, cross section and detector mass
results in a total of $N_{\mu^-}$ $\nu_\mu$ visible interactions,
registered by the apparatus, for both the conventional and
neutrino factory beams. In addition, in the case of conventional
beams, there will be $N_{e^-}$ $\nu_e$ visible interactions, due
to the intrinsic $\nu_e$ contamination, absent in the muon-induced
beam. If one is looking for $\nu_\mu\rightarrow \nu_e$
oscillations, then, in the neutrino factory\footnote{in reality at
the neutrino factory one measures the transitions $\nu_e
\rightarrow \nu_\mu$, since in a massive detector one can measure
much more easily the charge of muons than the charge of
electrons}, the sensitivity goes as:
\begin{equation}\label{eq:sensi-nufact}
  P(\nu_\mu\rightarrow \nu_e) \propto \frac{1}{N_{\mu}}
\end{equation}
since there is no $\nu_e$ contamination, while, in the case of a
conventional beam, one has:
\begin{equation}\label{eq:sensi-conv}
P(\nu_\mu\rightarrow \nu_e) \propto \frac{\sqrt{N_{e}}}{N_{\mu}}
\end{equation}
so, that if the $\nu_e$ contamination is a fraction $f$ of the
primary $\nu_\mu$ beam (assuming for simplicity identical $\nu_e$
and $\nu_\mu$ cross sections) we have:
\begin{equation}\label{eq:sensi-cc}
P(\nu_\mu\rightarrow \nu_e) \propto \frac{g}{\sqrt{N_{\mu}}}
\end{equation}
where $g = \sqrt{f}$. Although $g$ is a small quantity, the key
difference between conventional and muon-induced beams is clear
comparing equations \ref{eq:sensi-nufact} and \ref{eq:sensi-cc}.
In the first case the sensitivity improves {\it linearly} while in
the second improves only with the square root of the total
collected statistics.

Another issue concerns systematics in beam composition. While the
neutrino spectra from muon decay can be computed to a great
precision, the convoluted spectra in a conventional beam are
affected by a number of uncertainties, the most important of which
is the initial $\pi/K$ ratio in the hadron beam, which affects the
composition of the \numu:\nue:\nubarmu:\nubare~ beam. Typically,
these and other uncertainties translate into a systematic error at
the level of few per cent in the conventional neutrino fluxes, to
be compared with a few per mil, in the case of a neutrino factory.

Other important aspects to be considered when designing a
conventional beam are whether one prefers a wide or narrow band
beam and the energy regime. Beam energies range typically from few
hundred MeV to few hundred GeV, depending of the colliding hadron
beam and beam optics. High energy yields more interactions,
sufficiently low energy, we argue, a better control over
backgrounds and less beam uncertainties. We refer again
to\cite{barger1,barger2} for comparison of various energy regimes.

\section{The SPL neutrino beam}
\label{sec:spl}

\begin{table} [hbt]
\centering
\begin{tabular}{|l|c|}\hline
Mean beam power& 4MW\\\hline Kinetic energy&2.2 GeV\\\hline
Repetition rate& 75Hz\\\hline Pulse duration& 2.2 ms\\\hline
Number of protons per pulse (per second)& 1.5 $10^{14}$(1.1
$10^{16}$)\\\hline Mean current during a pulse&11 mA\\\hline
Overall lenght& 799 m\\\hline Bunch frequency (minimum time
between bunches)&352.2 MHz (2.84 ns)\\\hline
\end{tabular}
\caption{Basic SPL characteristics. \label{tab:spltab}}
\end{table}

\begin{figure}[htb]
\centerline{\epsfig{file=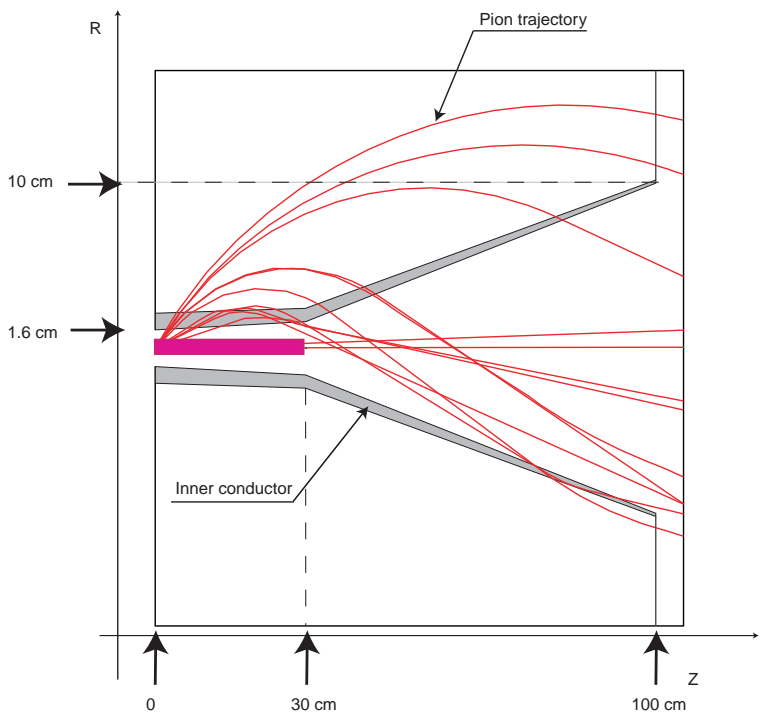,width=10.0cm}}
\caption{The horn focusing system.} \label{fig:horn}
\end{figure}

The planned Super Proton Linac is a proton beam of 4 MW power
which will be used as a first stage of the Neutrino Factory
complex. Its basic parameters are reported in Table
\ref{tab:spltab}. Pions are produced by the interactions of the
2.2 GeV proton beam with a liquid mercury target and focused (or
defocused, depending on the sign) with a magnetic horn (see Figure
\ref{fig:horn}). Next they transverse a cylindrical decay tunnel
of 1 meter radius and 20 meters length (found to be the optimal
decay length in\cite{donega}). We have used the {\tt MARS}
program\cite{mars} to generate and track pions, then analytical
calculations, described in\cite{donega} to compute the probability
that the neutrinos produced in both muon and pion decay reach a
detector of transverse dimensions $A$ located at a distance $L$
from the target.

\begin{figure}[htb]
\centerline{\epsfig{file=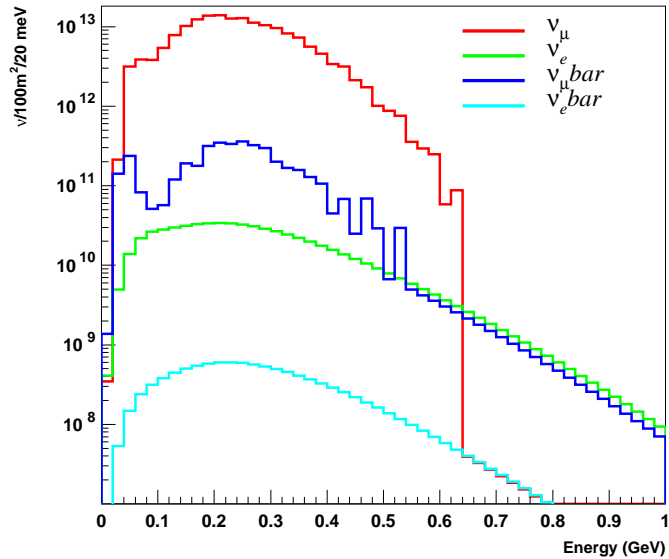,width=10.0cm}}
 \caption{The
SPL neutrino spectra, for $\pi^+$ focused in the horn. The fluxes
are computed at 50 km from the target, then scaled to the relevant
distances.} \label{fig:spl-spectra}
\end{figure}

The resulting neutrino spectra is shown in Figure
\ref{fig:spl-spectra}. Notice that the average energy of the
neutrinos is around 250 MeV and that the \nue~contamination of the
beam is at the level of few per mil. Due to the low energy of
protons, kaon production is strongly suppressed, resulting in both
less \nue~ contamination and better controlled beam systematics.

\section{Detector scenarios}
\label{sec:det}

\begin{figure}[htb]
\centerline{\epsfig{file=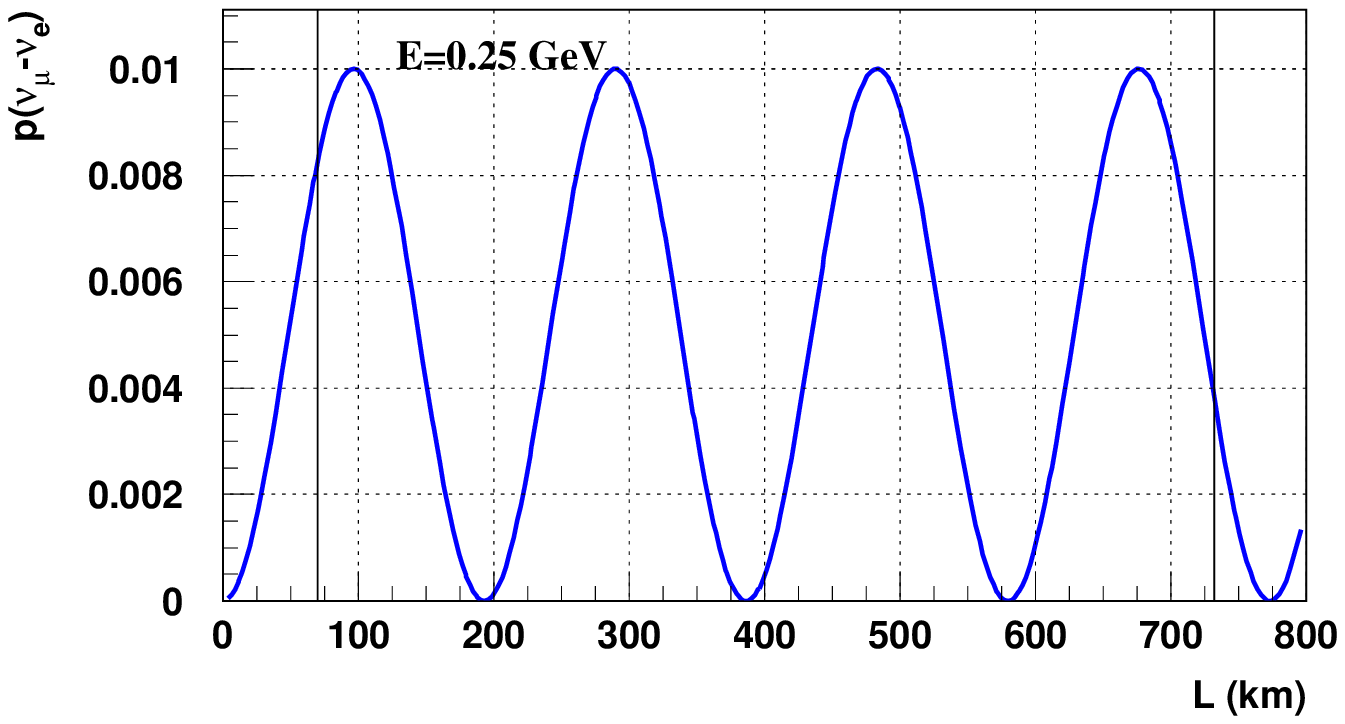,width=10.0cm}} \caption{The
oscillation probability \pnumunue, showing a first maximum around
100 km.} \label{fig:osc}
\end{figure}

 Figure \ref{fig:osc} shows the oscillation probability
\pnumunue as a function of the distance (for $\delta m_{23} = 3
\cdot 10^{-3}~$eV$^2$ \thetatt = 45$^\circ$). Notice that the
first maximum of the oscillation is at 100 km. Detection of
low-energy neutrinos at O(100 km) from the source requires a
massive target with high efficiency. Moreover, a search for
$\nu_e$ appearance demands excellent rejection of physics
backgrounds, namely $\mu$ mis-identification and neutral current
$\pi^0$ production, which should be controlled to a lower level
than the irreducible beam-induced background.

In this paper we consider two detector technologies, which have
demonstrated excellent performance in the low energy regime, while
being able to provide massive targets. These are, water Cherenkov
detectors, which have been developed by experiments such as
IMB\cite{Becker-Szendy:1993mb}, Kamiokande\cite{Hirata:1992ku},
and Super-Kamiokande\cite{Fukuda:1998tw}, and diluted liquid
scintillator detectors. This type of detectors were used by the
LSND experiment\cite{LSND:dif} and are being planned for the
forthcoming MiniBoone experiment\cite{Boone}, where both Cherenkov
and scintillation light is measured.

In spite of the fact that liquid scintillator apparatus provide, a
priory, more handles to reject backgrounds than their water
Cherenkov counterparts, the only truly massive detectors built so
far are of the latest type (compare Super-Kamiokande 50 ktons with
the sparse 499 tons of MiniBoone). For the water detector we have
conducted an extensive simulation followed by a full data
analysis. Instead, for the liquid scintillating detector, we have
worked out an educated guess, extrapolating published data, mainly
from LSND and MiniBoone. It is remarkable, however, the good level
of agreement that both approaches yield, as will be shown in the
remaining of the section.

As a base line we have considered 130 km, which is near the
maximum of the oscillation and equals the distance between CERN
and the Modane laboratory in the FREJUS tunnel, where one could
conceivably locate a large neutrino detector\cite{mosca,spiro}.

\subsection{Water Cherenkov detectors}

We have considered an apparatus of 40 kton fiducial mass and
sensitivity identical to the Super-Kamiokande experiment. The
response of the detector to the neutrino beams discussed in
section 3 was studied using the NUANCE\cite{casper} neutrino
physics generator and detector simulation and reconstruction
algorithms developed for the Super-Kamiokande atmospheric neutrino
analysis. These algorithms, and their agreement with real neutrino
data, have been described
elsewhere\cite{Fukuda:1998tw,Messier:1999kj,Shiozawa:1999sd}.

In the absence of neutrino oscillations, the dominant reaction
induced  by the beam is $\nu_{\mu}$ quasi-elastic scattering,
leading to a single observed (prompt) muon ring.  Recoiling
protons are well below Cherenkov threshold at the energies
discussed here, and hence produce no rings. To unambiguously
identify a potentially small $\nu_e$ appearance signal, it is
essential to avoid confusion of muons with electrons. Thanks to
the low energy of the SPL and its neutrino beam, the Cherenkov
threshold itself helps separate muons and electrons, since a muon
produced near the peak of the spectrum ($\sim 300 \,\hbox{MeV/c}$)
cannot be confused with an electron of comparable momentum;
instead it will appear to be a much lower-energy ($\sim
100\,\hbox{MeV/c}$) electron.

Particle identification exploits the difference in the Cherenkov
patterns  produced by showering (``e-like") and non-showering
(``$\mu$-like") particles. Besides, for the energies of interest
in this beam, the difference in Cherenkov opening angle between an
electron and a muon can also be exploited. Furthermore, muons
which stop and decay (100\% of $\mu^+$ and 78\% of $\mu^-$)
produce a detectable delayed electron signature which can be used
as an additional handle for background rejection.

\begin{figure}[htb]
\centerline{\epsfig{file=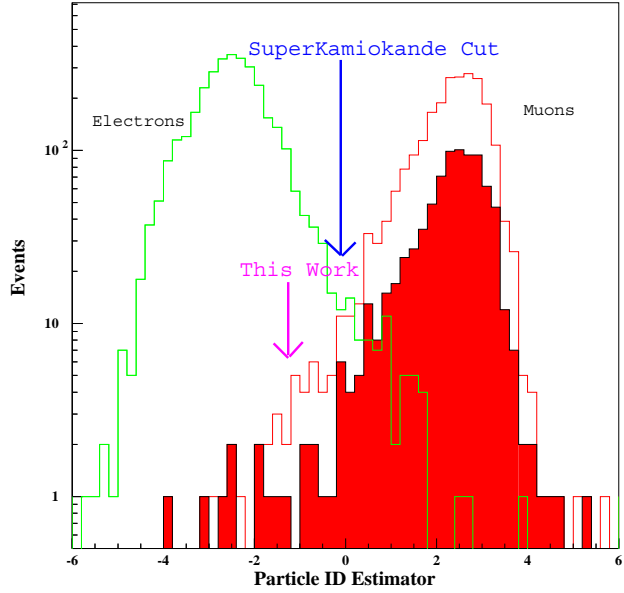,width=10.0cm}}
\vspace{-1.6cm} \caption{Rejection of $\nu_{\mu}^{CC}$ background
in a water Cherenkov detector. The particle ID estimator P (in
arbitrary units) is shown for the electron-like signal (left) and
muon-like background (right). The cut is set at -1, reducing
miss-identification of muons considerably at a negligible cost in
signal efficiency. Since most $\nu_{\mu}^{CC}$ events are followed
by a muon-decay signature, the background is further reduced by
accepting only events without a delayed coincidence (shaded
histogram on right).} \label{fig:pid}
\end{figure}

For this study, we have used the Super-Kamiokande particle
identification criteria, which are based on a maximum likelihood
fit of both $\mu$-like and e-like hypotheses. In terms of the
particle identification estimator P, shown in
Figure~\ref{fig:pid}, an event is e-like if $P_e > P_{\mu}+1$.
This cut introduces only a small inefficiency for true $\nu_e$
charged-current interactions, while reducing the $\nu_{\mu}$
background considerably. In addition, any event with an identified
muon decay signature is rejected from the e-like ($\nu_e$
appearance) sample.

Production of $\pi^0$ through neutral current resonance-mediated
and coherent processes is another major source of background,
which is, however, suppressed by the low energy of the beam and
the relatively small boost of the resulting $\pi^0$. This results
in events where the two rings are easily found by an standard
$\pi^0$ search algorithm, \'a la Super-Kamiokande. However, for
the events in which only a single ring is found we further apply
an algorithm\cite{Barszczak:thesis}, specially tailored to search
for low-energy $\gamma$'s (typically produced by asymmetric
decays). The algorithm {\it always} identifies a candidate for a
second ring, which, if the primary ring is truly a single
electron, is typically either very low energy, or extremely
forward.  If, on the other hand, two $\gamma$ from $\pi^0$ decay
are present, the second ring-candidate is usually the $\pi^0$
daughter which was missed by the standard pattern-recognition. By
requiring that the invariant mass formed by the primary ring and
the secondary ring-candidate is less than $45 \,
\hbox{MeV/$c^2$}$, almost all remaining $\pi^0$ contamination of
the single-ring, e-like sample is removed.

\begin{figure}[htb]
\vspace{-0.5cm} \centerline{\epsfig{file=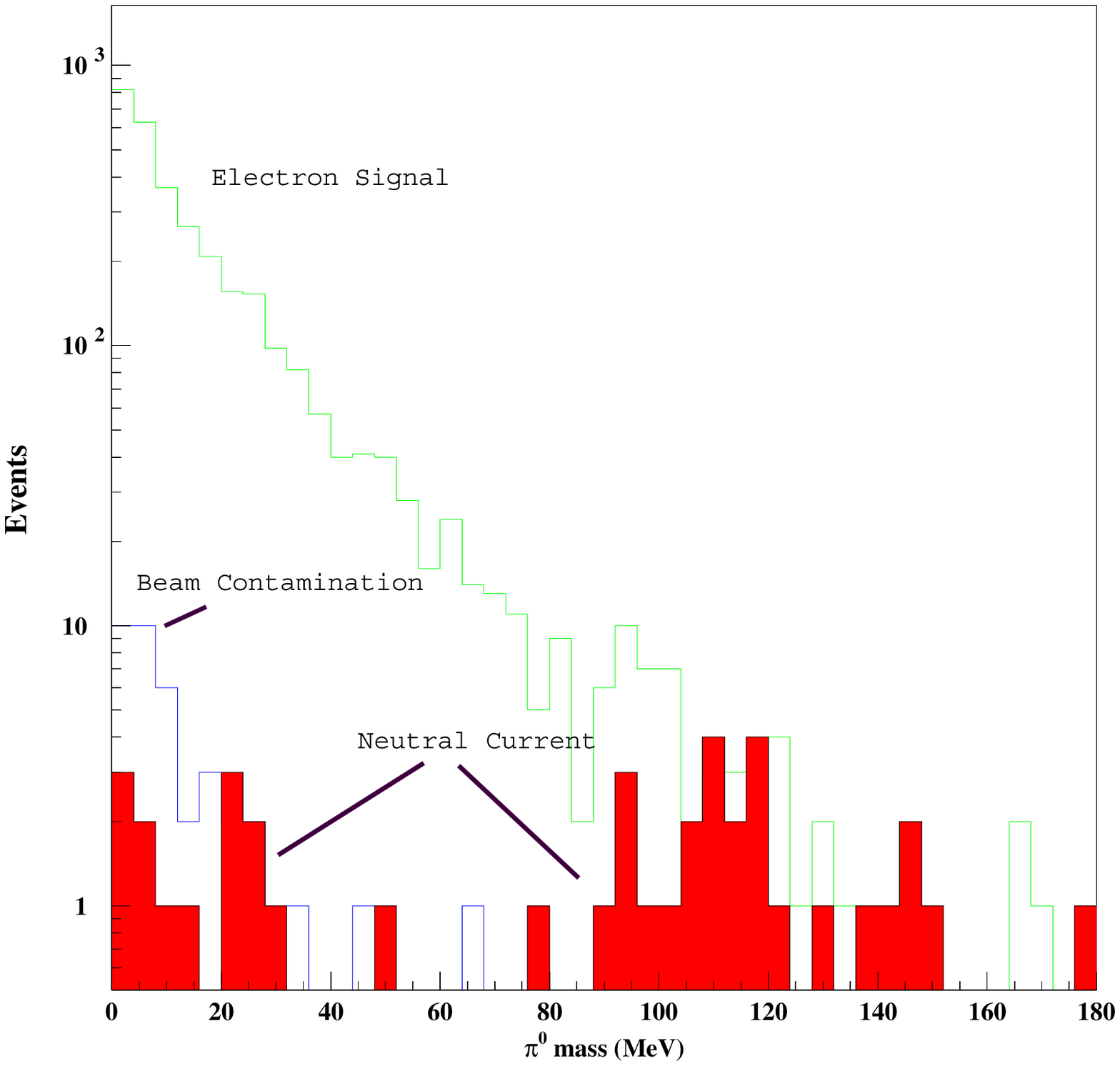,width=10.0cm}}
\vspace{-1.6cm} \caption{Rejection of $\pi^0$ background in a
water Cherenkov detector. To reject $\pi^0$ in which a weak second
ring was missed, each showering event with a single identified
ring is analyzed to find the most likely direction and energy of
an additional ring. The invariant mass formed by the second
ring-candidate and the original ring tends to zero for true
electrons (unfilled histograms), but is O($m_{\pi^0}$) for many
neutral-current background events. (shaded).} \label{fig:pi0}
\end{figure}

The background in each category ($\nu_{\mu}$ charged-currents,
$\nu_e$ contamination in the beam, and neutral currents) remaining
after all selections, and the efficiency for signal, after each
cut is summarized in Table~\ref{tbl:Events} for the $\pi^+$
focused beam and Table~\ref{tbl:antiEvents} for the $\pi^-$
focused beam.  Contamination by $\nu_e$ from muon decay in the
secondary beam is dominant.

\begin{table}
\begin{tabular}{|c|c|c|c|c|c|c|}
\hline
 &&  & Fit in fiducial volume & Tight & &  \\
 & Initial& Visible & Single-ring & particle & No & \\
Channel & sample & events & $100 - 450 \, \hbox{MeV/$c^2$}$ & ID &
$\mu
\rightarrow e$ & $m_{\gamma \gamma} < 45 \, \hbox{MeV/$c^2$}$ \\
\hline
$\nu_{\mu}^{CC}$ & 3250& 887 & 578.4 & 5.5 & 2.5 & 1.5 \\
$\nu_e^{CC}$& 18&  12. & 8.2 & 8.0 & 8.0 & 7.8 \\
NC & 2887& 36.9 & 8.7 & 7.7 & 7.7 & 1.7 \\ \hline $\nu_{\mu}
\rightarrow \nu_{e}$ & \- & 82.4\% & 77.2\% & 76.5\% & 70.7\%\\
\hline
\end{tabular}
\caption{Summary of simulated data samples a $\pi^+$  focused
neutrino beam.  The first three lines show the expected background
surviving the selection at each stage for a 5-year exposure of a
40 kton (fiducial) water detector located at 130 km from the
source. The bottom line shows the efficiencies for the $\nu_{\mu}
\rightarrow \nu_e$ signal.  The numbers in the rightmost column
(after all cuts) represent the sample used to estimate the
oscillation sensitivity.} \label{tbl:Events}
\end{table}

\begin{table}
\begin{tabular}{|c|c|c|c|c|c|c|}
\hline
 &&  & Fit in fiducial volume & Tight & &  \\
 & Initial& Visible & Single-ring & particle & No & \\
Channel & sample & events & $100 - 450 \, \hbox{MeV/$c^2$}$ & ID &
$\mu
\rightarrow e$ & $m_{\gamma \gamma} < 45 \, \hbox{MeV/$c^2$}$ \\
\hline
$\stackrel{-}{\nu}_{\mu}^{CC}$ & 539& 186 & 123 & 2.3 & 0.7 & 0.7 \\
$\stackrel{-}{\nu}_e^{CC}$ & 4& 3.3 & 3 & 2.7 & 2.7 & 2.7 \\
NC & 687& 11.7 & 3.3 & 3 & 3 & 0.3 \\ \hline
 $\stackrel{-}{\nu}_{\mu}
\rightarrow \stackrel{-}{\nu}_{e}$ & \- & 79.3\% & 74.1\% & 74.0\%
& 67.1\% \\ \hline
\end{tabular}
\caption{Summary of simulated data samples a $\pi^-$ focused
neutrino beam. The first three lines show the expected background
surviving the selection at each stage for a 5-year exposure of a
40 kton (fiducial) water detector located at 130 km from the
source. The bottom line shows the efficiencies for the
$\stackrel{-}{\nu}_{\mu} \rightarrow \stackrel{-}{\nu_e}$ signal.
The numbers in the rightmost column (after all cuts) represent the
sample used to estimate the oscillation sensitivity.}
\label{tbl:antiEvents}
\end{table}

\subsection{Liquid scintillator detectors}

Liquid scintillator technology has been used by the LSND
experiments\cite{LSND:dif} to detect an small amount of low energy
\nue~ events in an intense \numu~ beam. The very same technique
will be used by the forthcoming MiniBoone experiment\cite{Boone}.

In diluted liquid scintillator detectors both Cherenkov and
scintillation lights are measured. They can be separated given the
different light emission timing and direction. The Cherenkov light
pattern can be used to separate $\pi^0$ and $\mu$  from electrons
while
 the ratio between scintillation and Cherenkov
light provides additional handles to separate muons from
electrons.The energy range and the rejections against background
needed for those experiments nicely match the requirements of our
study as summarized in Table \ref{tab:bck}.

The obvious shortcut of the liquid oil technology is its relative
high price compared with water. Indeed, the mass of the largest
liquid oil neutrino detector (the forthcoming MiniBoone) is two
orders of magnitud smaller than the mass of the largest water
neutrino detector, Super-Kamiokande. One could hardly afford truly
large, 50 kton or more liquid oil detectors.. Nevertheless, for
the sake of a fair comparison between both technologies, in the
following we will assume a detector identical to MiniBoone (449
ton of pure mineral oil, fiducial is 382 ton, with a photocathode
 surface coverage of 10\%) but inflated to a 40 kton fiducial detector.

\begin{table}[htb]
\begin{center}
\begin{tabular}{|c|c|}
\hline
Reaction    &   Suppression factor\\
\hline
$\nu_\mu C \rightarrow \mu^- X$         & $10^{-3}$\\
$\nu_\mu C \rightarrow \nu_\mu\pi^\circ X$  & $10^{-2}$\\
$\nu_\mu C \rightarrow \mu^-\pi X$  & $10^{-4}$\\
$\nu_\mu C \rightarrow \nu_\mu\pi X$    & $10^{-3}$\\
$\nu_\mu e \rightarrow \nu_\mu e$   & $10^{-1}$\\
\hline
$\nu_e C \rightarrow e^- X$     & 0.5\\
\hline
\end{tabular}
\caption{Background suppression and signal efficiency in the
MiniBooNE detector. Numbers are quoted in the 50 MeV - 1 GeV
energy range.} \label{tab:bck}
\end{center}
\end{table}

Neutrino-$^{12}C$ cross sections are taken from
reference\cite{RPA}. They come from an upgraded version of the
continuous random phase approximation method used to compute
$\nu-^{12}C$ cross-sections and in average they are lower by about
$\sim 15\%$ from what quoted by the MiniBoone experiment.

Table \ref{tab:events} shows the background event distributions,
assuming no \numu-\nue oscillation (e.g., driven by \thetaot), for
a 200 kton-year exposure to a $\pi^+$ and a $\pi^-$ focused beams.
As before, intrinsic \nue~ contamination from the beam results to
be the dominant background.

\begin{table}[htb]
\begin{center}
\begin{tabular}{|c|c|c||c|c|c|}
\hline \multicolumn{3}{|c|}{$\pi^+$ focused beam} &
\multicolumn{3}{|c|}{$\pi^-$ focused beam}\\
\hline
Channel & Initial sample & Final sample & Channel & Initial sample & Final sample \\
\hline
$\numu^{CC}$ & 2538 & 2.5 & $\nubarmu^{CC}$ & 451 & 0.5 \\
$\nue^{CC}$ & 12 & 6 & $\nubare^{CC}$ & 2.3 & 1.0\\
NC (visible) & 48 & 0.5 & NC & 10 & $<0.1$ \\
\hline
\numunue & 100\% & 50\% & $\nubarmu \rightarrow \nubare$ & 100\% & 50 \%  \\
\hline
\end{tabular}
\caption{Summary of data samples in a $\pi^+$ and in a $\pi^-$
focused neutrino beam. Numbers refer to a  liquid scintillator
detector of 40 kton located at a distance of 130 km from the
source and a run of 5 years.} \label{tab:events}
\end{center}
\end{table}

As one can see comparing tables
\ref{tbl:Events},\ref{tbl:antiEvents} and \ref{tab:events}, our
estimations for the rates and performance of the liquid oil
detector match quite well with our calculations concerning the
water detector. The performance of both devices is quite similar,
although the liquid scintillator is able to reject more neutral
currents than the water Cherenkov (as one expects, given the extra
handle provided by the scintillation light). The dominant
background in both cases is the beam \nue~ contamination. The
conclusion is that one would probably prefer, for this experiment,
a water detector, where one can afford truly gargantuan sizes.

\section{Sensitivity}
In this section we illustrate the sensitivity that a 40 kton water
or liquid oil detector, located at 130 km from the source would
have to the various parameters of the neutrino CKM matrix.

\subsection{Sensitivity to the atmospheric parameters}
A 40 kton  detector placed at L=130 Km has excellent opportunities
of precision measurements of $\sin^2\thetatt$ and $\Delta
m^2_{23}$ with a \numu disappearance experiment. Given the mean
beam energy of the \numu beam
 $(1.27 \cdot L/E)^{-1}=1.6 \cdot 10^{-3} \;\rm{eV^2/c^4}$ and so
$p(\numu\rightarrow\numu)$ results to be just at its minimum.

\begin{figure}[htb]
\vspace{-0.2cm}
\centerline{\epsfig{file=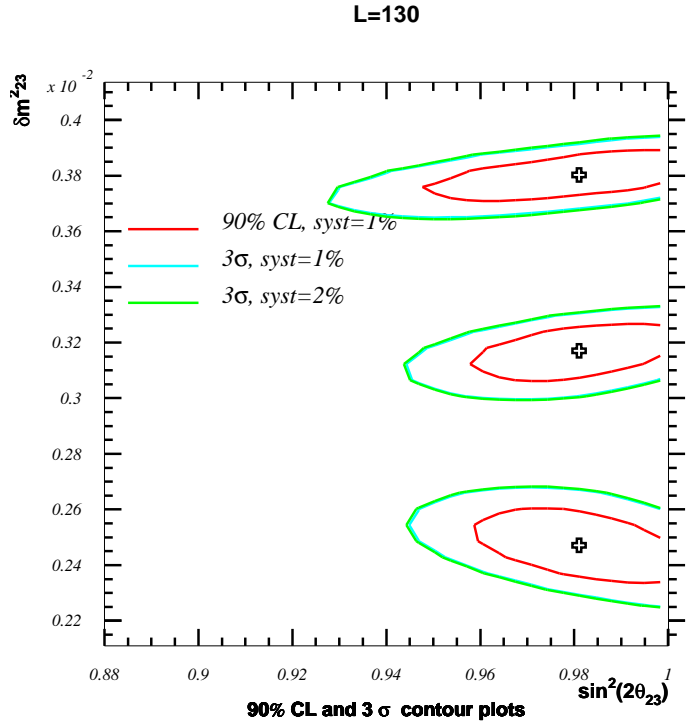,width=10cm}}
\vspace{-1.6cm} \caption{Fits of \dmatm~ (eV$^2$),\sthetatt~ plane
after 5 years of run, for a systematic error of  2\% and a
distance of 130 km. The crosses sign the initial points
$(0.98,3.8\cdot10^{-3})$, $(0.98,3.2\cdot10^{-3})$,
$(0.98,2.5\cdot10^{-3})$ in \dmatm,\sthetatt~ coordinates.}
\label{fig:result130}
\end{figure}

To illustrate the precision in measuring \dmatm~ and \thetatt~ in
case of positive signal Figure \ref{fig:result130} shows the
result of a 200 kton-years exposure experiment (5 years of a 40
kton detector)  in case the oscillation occurs with \sthetatt=0.98
and $\dmatm=3.8, 3.2$ or 2.5 $eV^2/c^4$. The computation is
performed defining 4 energy bins in the 0.1-0.7 GeV energy range
and including Fermi motion, that is by far the most limiting
factor to energy reconstruction at these energies. See\cite{mauro}
for more details. We find that $\Delta m^2_{23}$ can be measured
with a standard deviation of  $1\cdot10^{-4}\quad {\rm eV^2/c^4}$
while $\sin^2 2\thetatt$ is measured at the 1\% level.

\subsection{Sensitivity to \thetaot in the SMS-MSW scenario}
Here we assume, for simplicity that the solar parameters, $\delta
m_{12}$ and $\theta_{12}$ correspond to the small angle solution
of the solar neutrino problem. In this case, the oscillation
probability simplifies to:
\begin{equation}\label{probs}
  P_{\nu_e\nu_\mu}  =  \sin^2 2 \;\theta_{13}  \sin^2
\theta_{23}
 \sin^2 \frac{\Delta m^2_{23} L}{4 E_\nu}
\end{equation}
that is, the oscillation depends only on the atmospheric
parameters and $\theta_{13}$. For the present study, only
statistical errors are considered. Given the 2.5:1 disparity
between expected beam and detector backgrounds, it is likely that
beam-related uncertainties will be the most important, and these
can be controlled by measuring the beam with a near detector and
using data from the HARP\cite{Catanesi:1999xr} experiment to
refine the hadronic production model.

\begin{figure}[htb]
\centering \epsfig{figure=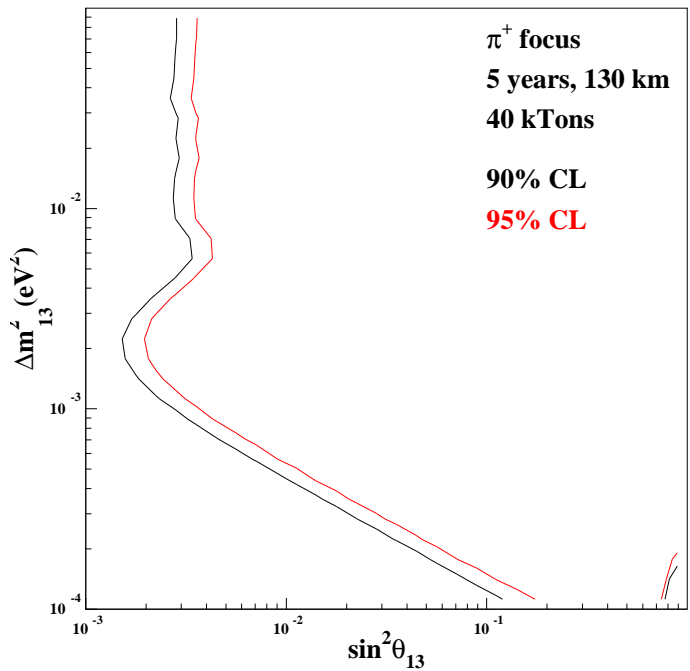,width=10.0cm}
 \caption{Oscillation sensitivity for $\pi^+$
focused neutrino beams.  The outer(inner) contours are the regions
where the expected confidence level to reject the oscillation
hypothesis in the absence of oscillation exceeds 90\%(95\%).}
\label{fig:sPlus}
\end{figure}

As an example Figure~\ref{fig:sPlus} shows the expected
sensitivity for a 5-year run with a 40 kton (fiducial) water
target at a distance of 130 km, using a $\pi^+$ focused neutrino
beam from the SPL.

\subsection{Sensitivity to CP in the LMA-MSW scenario}
In the remaining of this section we will assume that the solar
parameter lie in the upper range of the large mixing angle
solution (LMA-MSW) of the solar problem, specifically we will
assume maximal mixing in the solar sector and $\delta m_{12}^2 =
10^{-4}$~eV$^2$. We consider a water detector.

Unfortunately, the $\stackrel{-}{\nu} + {}^{16}O$ cross-section is
approximately six times less than that for $\nu + {}^{16}O$ at
these energies, diminishing the experiment's sensitivity to
CP-violation considerably (about the same considerations apply to
Carbon, in the case of liquid oil detectors).

We follow the approach in\cite{golden,jordi} and fit
simultaneously the CP phase $\delta$ and \thetaot. Notice that,
although we apply a full three family treatment to our
calculations, including matter effects, these are not important at
the short distances and low energies considered. Notice also that
the measurement of the solar parameters will be performed by
Kamland\cite{kamland}, well before the experiment described here,
and that the determination of the atmospheric parameters, done
with muon disappearance, as illustrated above, is also largely
uncorrelated from the measurement of the other parameters.

\begin{figure}[htb]
\centering \epsfig{figure=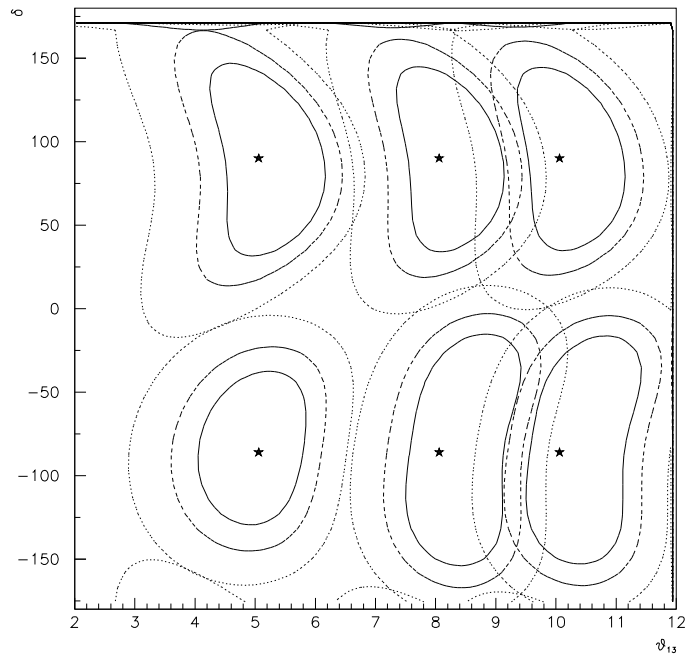,width=10.0cm} \caption{one
sigma, 90 \% and 99 \% confidence level intervals resulting from a
simultaneous fit to the \thetaot~and $\delta$ parameters. The
generated values were $\thetaot=5^\circ,8^\circ,10^\circ$,
$\delta=\pm 90^\circ$. A full three family treatment is used.
Statistical errors as well as those due to background substraction
are taken into account. We have considered a 10 year antineutrino
and a 2 year neutrino run, at 130 km with a 40 kton detector. }
 \label{fitcp}
\end{figure}

Figure~\ref{fitcp} shows the confidence level contours for a
simultaneous fit of \thetaot and $\delta$, corresponding to three
values of \thetaot, $\thetaot=5^\circ,8^\circ,10^\circ$, and a
maximally violating CP phase, $\delta=\pm 90^\circ$. The results
include statistical errors as well as those due to background
subtraction. Since the sensitivity is dominated by the low
antineutrino statistics, we have considered for this exercise a 10
year run with focused $\pi^-$ and a 2 year run with focused
$\pi^+$.

Inspection of Figure \ref{fitcp} permits to draw two immediate
conclusions. The first one is that the sensitivity to CP does not
worsen very much when \thetaot~ becomes (moderately) smaller, as
pointed out in\cite{romanino,golden,jordi}. The second is that, at
90 \% confidence level, a maximally violating CP phase ($\delta =
\pm 90^\circ$) would be just distinguishable from a non CP
violating phase ($\delta = 0^\circ$). Recall that this is only in
the upper limit of the LMA regime. In conclusion, this experiment
would offer a chance to observe CP violation if nature would
conspire to offer a very lucky scenario (maximal CP violation,
solar square mass difference as large as allowed by current data).

\begin{figure}[htb]
\centering \epsfig{figure=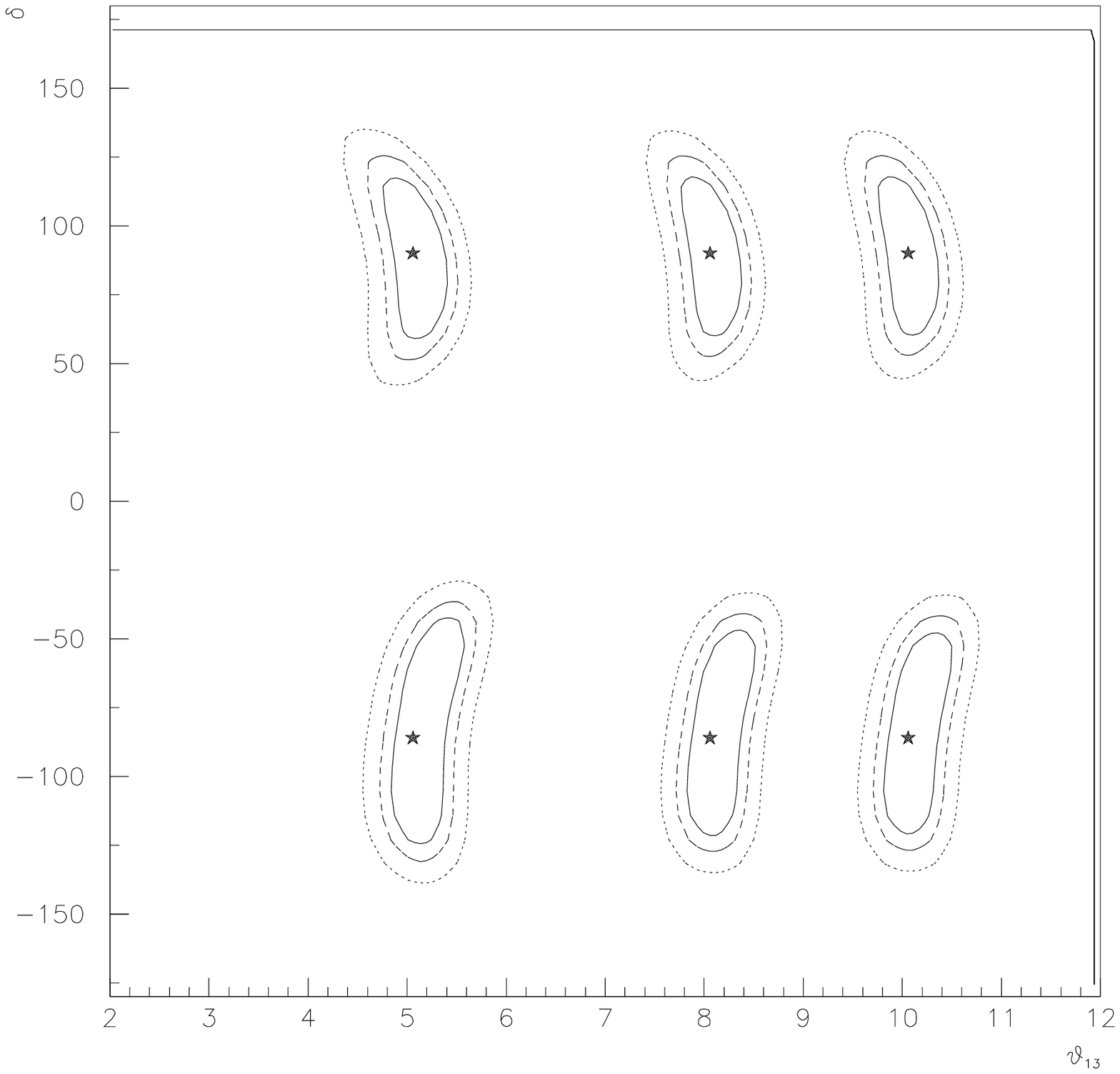,width=10.0cm}
 \caption{one sigma, 90 \% and 99 \% confidence
level intervals resulting from a simultaneous fit to the
\thetaot~and $\delta$ parameters. The generated values were
$\thetaot=5^\circ,8^\circ,10^\circ$, $\delta=\pm 90^\circ$. A full
three family treatment is used. Statistical errors as well as
those due to background substraction are taken into account. We
have considered a 10 year antineutrino and a 2 year neutrino run,
at 130 km with a 400 kton detector.} \label{fig:cpbig}
\end{figure}

Figure \ref{fig:cpbig} shows the result of the same fit, now
assuming a very large water detector, such as the proposed
UNO\cite{uno} water Cherenkov apparatus, with a fiducial mass of
400 ktons. Clearly, the prospects to observe CP violation are much
improved.

\section{Conclusions}
We have examined the physics potential of a low energy super beam
which could be produced by the CERN Super Proton Linac. Water
Cherenkov and liquid oil scintillator detectors have been
considered. Detailed calculations have been performed for the case
of the water detector.

The low energy of the beam studied has several advantages. Beam
systematics is reduced with respect to high energy, since one is
below kaon production. Furthermore, $e/\mu$ and $e/\pi^0$
separation in a water (liquid oil) detector is near optimal at
this low energies. The drawback are the small anti neutrino cross
sections, which are more than a factor five smaller than neutrino
cross sections.

The peak of the oscillation is at a distance of about 100 km. An
ideal location, at 130 km from CERN exists, the Modane laboratory
in the Frejus tunnel.

A ``moderate" size detector (``only" twice as big a
Super-Kamiokande) at this baseline could, in a five year run,
improve our knowledge of the atmospheric parameters by about one
order of magnitude (with respect to the expected precision of next
generation neutrino experiments, such as Minos). It could also
measure \thetaot~if its magnitude is bigger than about $3^\circ$,
again, more than one order of magnitude the precision of next
generation experiments. For comparison, one could do slightly
better in the neutrino factory (about a factor two to three) for
what concerns the atmospheric parameters, and more than one order
of magnitude better for \thetaot.

Such a detector could also, if the solution to the solar neutrino
problem lies in the upper part of the LMA, distinguish,
eventually, a maximally violating CP phase. Here, the performance
if much worst than the one expected for the neutrino factory. For
CP violation studies a very large detector, \'a la UNO (400 kton
fiducial mass) is mandatory.
\section{Acknowledgements}
We wish to thank the Super-Kamiokande collaboration for allowing
use of its simulation and analysis software in this study.

\end{document}